\documentclass[aip,jcp]{revtex4-1}
\usepackage{amsmath,amssymb}
\usepackage{graphicx,color}
\begin{document}
%\preprint{0.30}
\title{Structure, stability, and mobility of small Pd clusters
on the stoichiometric and defective TiO$_2$ (110) surfaces}
\author{Jin Zhang}
\author{Anastassia~N.~Alexandrova}
\email{ana@chem.ucla.edu}
\affiliation{Department of Chemistry and Biochemistry, University of California,
Los Angeles, Los Angeles, CA 90095-1569}
\date{\today}

\begin{abstract}
We report on the structure and adsorption properties of Pd$_n$ ($n=1-4$) clusters
supported on the rutile TiO$_2$ (110) surfaces with the possible presence
of a surface oxygen vacancy or a subsurface Ti-interstitial atom.
As predicted by the density functional theory,
small Pd clusters prefer to bind to the stoichiometric titania surface
or at sites near subsurface Ti-interstitial atoms.
The adsorption of Pd clusters changes the electronic structure of the underlying surface.
%and in particular, leads to surface reconstruction and possible additional pathways
%for migration of Ti-interstitial atoms to form TiO$_2$ islands on the surface.
For the surface with an oxygen vacancy,
the charge localization and ferromagnetic spin states are found to be largely attenuated
owing to the adsorption of Pd clusters.
The potential energy surfaces of the Pd monomer on different types of surfaces are also reported.
The process of sintering is then simulated via the Metropolis Monte Carlo method.
The presence of oxygen vacancy likely leads to the dissociation of Pd clusters.
On the stoichiometric surface or surface with Ti-interstitial atom,
the Pd monomers tend to sinter into larger clusters,
whereas the Pd dimer, trimer and tetramer appear to be relatively stable
below $600$ K.
This result agrees with the standard sintering model of transition metal clusters
and experimental observations.

\end{abstract}
\maketitle

\section{Introduction}\label{sec:intro}
In the recent decades, the heterogeneous catalysis of metal/metal-oxides
has advanced to the sub-nano regime,
where one promising catalyst architecture is metal clusters
composed of only a few atoms and supported by oxide surfaces.
The chemical properties of these sub-nano metal clusters have been shown
to be very different from those of corresponding bulks or
nanoparticles with hundreds of atoms.
In model studies of chemical reactions catalyzed by these systems,
the oxides surface does not only provide the support for the cluster
but also directly participates in the reaction by supplying charges,
adsorption sites, and additional reaction paths.\cite{HenrichCox1996,Henry1998}
As one of the most popular model systems on the transition metal-oxide surfaces,
the rutile TiO$_2$ (110) provides rich and controllable surface conditions
for studies on various catalytic reactions.\cite{Diebold2003}

Experimentally, though the TiO$_2$ surface is often prepared in ultra high vacuum (UHV)
conditions to maintain the cleanliness of the surface as much as possible,
it is known that many TiO$_2$ surfaces prepared by ion sputtering and annealing
in vacuum have localized band-gap states with Ti$3d$
character.\cite{Zhang1991,Heise1992,Diebold1994,Henderson1998,Diebold2003}
These gap states, which lie about $0.8$--$1.1$ eV
below the conduction band minimum (CBM),
are believed to be involved with imperfections of the surface,
such as surface oxygen vacancies\cite{Thornton2010,Thornton2010PNAS,Hammer2011}
or near-surface Ti-interstitial atoms.\cite{Wendt2008,Lira2011}
What roles these defects play in the metal/metal-oxides catalysis
and how they interact with the metal clusters is not well-understood.

On the scale of several atoms,
the size of the metal clusters can have a dramatic effect on the catalytic
activity.\cite{Landman2001,Anderson2004,Bernd2006,Harding2008,Anderson2010,Bonanni2011}
For example,
in a recent temperature-programmed reaction experiment done by the Anderson group,
the carbon monoxide oxidation rate in \textit{vacuo} shows
a $150$\% jump when the size of the deposited Pd$_n$ cluster increases from monomer (Pd$_1$)
to dimer (Pd$_2$).\cite{Kaden2009}
In the meantime,
it is well-known that metal clusters are vulnerable to sintering processes,
in which small sized clusters gradually group into larger ones
and finally lose their unique catalytic activity.\cite{Goodman2003,Wallace2005,Zhang2008,Celik2010}

A good understanding of these intriguing catalytic behaviors of Pd clusters on titania
requires an extensive microscopic knowledge of how Pd clusters bind and move
on the titania surfaces with the possible existence of surface defects.
A theoretical study, with controlled surface cleanliness,
particle size, and number density, is ideal for this purpose.
In this paper,
based on the density functional theory (DFT) with Hubbard $U$ correction,
we present a thorough study on the adsorption geometries, energetics, electronic structures,
and magnetic properties of Pd$_n$ ($n=1-4$) clusters
as well as the potential energy surface (PES) of the Pd monomer
on the stoichiometric and defective titania surfaces.
With these data we perform Monte Carlo (MC) simulations and further investigate
the sintering patterns of Pd clusters under experimentally relevant conditions.

\section{Method of Calculation}\label{sec:method}
The calculations were performed at the DFT level
with a plane-wave basis set and ultrasoft pseudopotentials
implemented in \textit{Quantum Espresso}.\cite{GiannozziBaroni2009}
The Ti 3s- and 3p-states were treated as valence states.
The number of valence electrons considered for Ti, Pd, O and H
were 12, 10, 6 and 1 respectively.
The cutoff energy for the plane-wave basis set was chosen to be $32$ Ry ($\approx 435$ eV).
When the surface oxygen vacancy or Ti-interstitials are present,
the additional electrons from these defects may localize on nearby Ti atoms and,
reduce these Ti$^{4+}$ ions to Ti$^{3+}$.\cite{}
These highly localized states are known to be poorly described by traditional
density functionals, including the local density approximation
and the generalized gradient approximation (GGA).\cite{}
To describe the exchange-correlation effect of valence electrons,
we adopted the DFT+$U$ approach (implemented in the simplified Dudarev\cite{Dudarev1998} scheme)
with the Perdew-Burke-Ernzerhof (PBE) functional.\cite{PBE1996}
The $U$ term was only applied to Ti atoms with its value being computed
self-consistently \cite{Cococcioni2005} as $4.2$ eV.
Spin-polarized calculations with fixed multiplicity were performed in all cases.
A gaussian smearing with width of $5$ mRy ($\approx 68$ meV) was applied to the occupation of electron bands.
The space charges were assigned to individual atoms through
the Bader charge analysis.\cite{Bader1994,Henkelman2006}

With above settings, the lattice constant of titania in equilibrium reads
$a=4.67$ {\AA} and $c=3.02$ {\AA},
which agrees with the experimental value ($a=4.58$ {\AA} and $c=2.95$ {\AA})
with errors smaller than $2$\%.
These values were used to construct a stoichiometric ($4\times2$) surface slab,
which was the starting point of all subsequent calculations with adsorbates.
One Pd cluster per supercell gives $1/8$ ML coverage,
where $1$ ML is equivalent to the density of 1 cluster per TiO$_2$ ($1\times1$) unit cell
($5.2\times10^{14}$ cm$^{-2}$).
The vacuum distance between two images of slabs is chosen to be $13$ {\AA},
contributing an error in energy of the order of $2$ meV.
The size of the supercell is large enough to use a single point (shifted from the $\Gamma$ point)
in the reciprocal space to evaluate the total energy.
The adsorbates were only placed at one side of the slab.
The energies of gas phase Pd$_n$ cluster were calculated
in a $15$ {\AA} cubic box at the $\Gamma$ point.
A good balance between the computation speed and accuracy can be achieved
using a four-trilayer of TiO$_2$ unit with the top two trilayers fully relaxed
and the remaining two constrained to their bulk equilibrium positions.
To avoid the band-gap narrowing and spurious surface state owing to this additional constraint,
we saturate the dangling bonds of Ti and O at the back of the slab
using pseudo-hydrogen atoms with fractional charges $4/3$ and $2/3$ respectively,
as suggested by Kowalski \textit{et~al.}\cite{Kowalski2009,Kowalski2010}
The total number of atoms including the pseudo-hydrogens per stoichiometric supercell is $208$.

\section{Results and discussion}\label{sec:results}
In this section, we show results on the adsorption properties of Pd$_n$ clusters ($n=1-4$)
on the stoichiometric and two defective TiO$_2$ (110) surfaces:
the surface with an oxygen vacancy and the one with a subsurface Ti-interstitial atom.
With the energetics obtained from DFT+$U$ calculations,
we perform MC simulations on the growth and sintering of small Pd clusters.

\begin{figure}[tb]
	\begin{center}
		\includegraphics[width=0.4\textwidth]{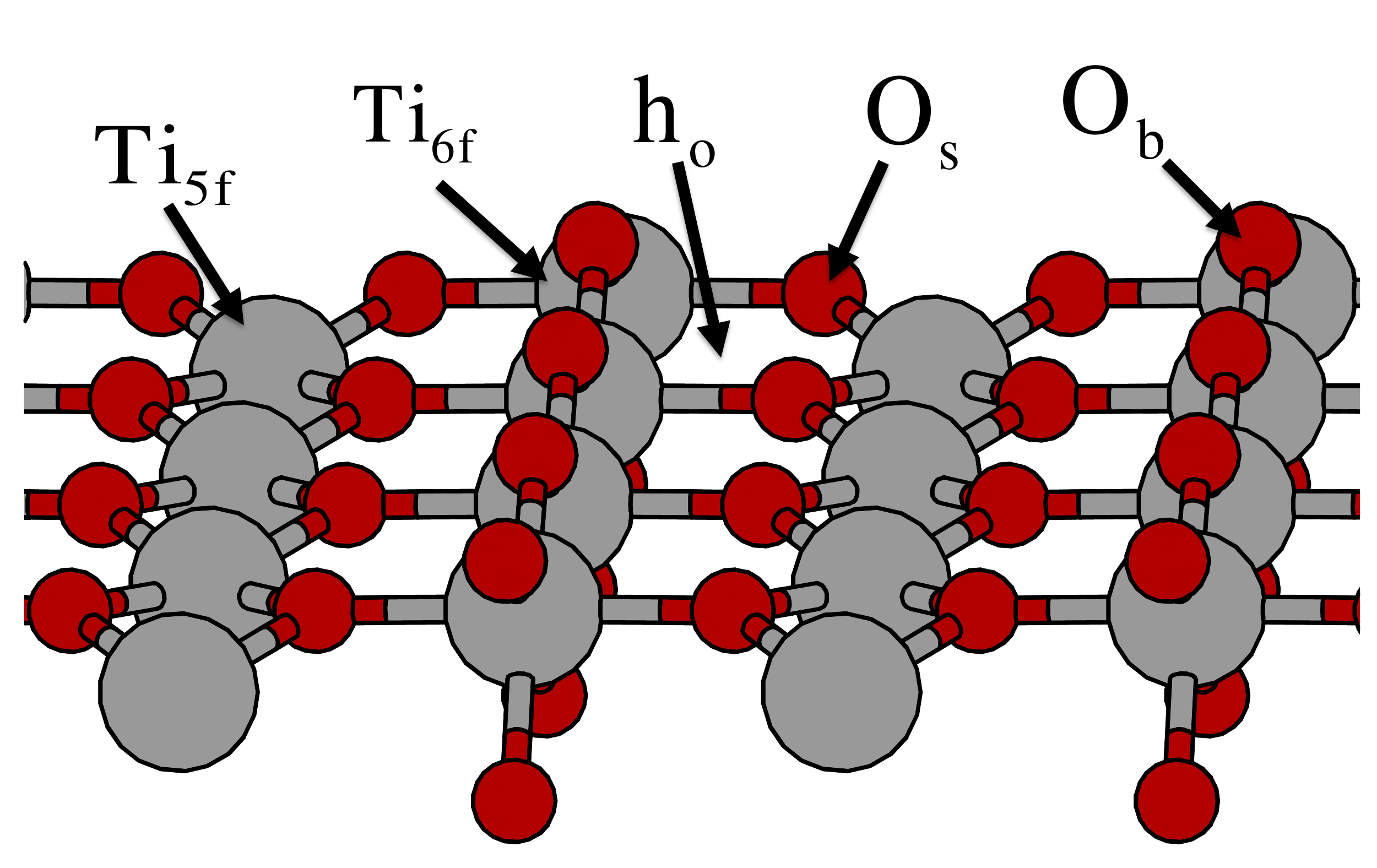}
	\end{center}
	\caption{The stoichiometric rutile TiO$_2$ (110) surface.
	Atoms are labeled as follows:
	Ti$_{5f}$ and Ti$_{6f}$ --- five- and six-fold coordinated Ti atom;
	O$_b$ --- surface bridging oxygen atom;
	O$_s$ --- first layer in-plane oxygen atom;
	h$_o$ --- hollow site between two O$_s$ atoms and a nearby O$_b$ atom.
	}
	\label{fig:notion}
\end{figure}

To facilitate the characterization of major adsorption sites,
we introduce the following notions (taking the stoichiometric surface as an example),
shown in Fig.~\ref{fig:notion}.
The five- and six-fold coordinated Ti atoms are marked as Ti$_{5f}$ and Ti$_{6f}$.
The two-fold coordinated surface oxygen atom in the bridging row is labeled as O$_b$,
whereas O$_s$ represents the three-fold oxygen atom in the first trilayer of the TiO$_2$ unit.
The hollow site between the two O$_s$ and a nearby O$_b$ atom is denoted as h$_o$.

\subsection{Gas phase Pd$_n$ ($n=2-4$) cluster}\label{subsec:gas}
The Pd atom has a $4d^{10}5s^0$ configuration.
The sp-d hybridization occurs when Pd atoms are brought into contact,
leading to partial occupancy in the $s$- and $d$- atomic orbital.
The exchange energy among these partially filled $4d$ states
tends to drive the system to higher multiplicity.
In bulk, though being paramagnetic,
Pd is very close to satisfy the Stoner criterion \cite{Stoner1938} and,
becomes ferromagnetic with a small lattice expansion.\cite{Moruzzi1989,Chen1989}

Several studies have been dedicated to the energetics, structures and magnetic properties of
Pd clusters in gas phase.\cite{Barreteau2000,Kumar2002,Futschek2005,Vega2006}
In the case of small Pd clusters ($n=2-4$),
the ferromagnetic configuration with magnetic moment $M=2$ $\mu_B$
was identified to be the global energy minimum.
The spin-singlet isomers are generally a few tenths of eV higher in energy.
In Tab.~\ref{tab:cluster},
we give the binding energy per atom: $E_b(n)=-[E(\mbox{Pd}_n)-nE(\mbox{Pd}_1)]/n$,
where $E(\mbox{X})$ is the total energy of the component X,
the dissociation energy: $E_d(n)=E(\mbox{Pd}_{n-1})+E(\mbox{Pd}_1)-E(\mbox{Pd}_n)$,
which gives the energy cost of removing one Pd atom from the Pd$_n$ cluster,
and the average Pd-Pd bond length $d$ for $n=2-4$.
The binding energies obtained by Futschek and co-workers\cite{Futschek2005} are also listed for comparison.

\begin{table}[tb]
\begin{tabular}{ccccccc}
\hline
n & Geometry    & $M$ ($\mu_B$) & $d$ ({\AA})    & $E_b$ (eV)
	& $E_b$ (eV) [Ref.~\onlinecite{Futschek2005}] & $E_d$ (eV) \\
\hline
2 & Dimer       & $2$ & $2.51$ & $0.634$ & $0.646$  & $0.634$ \\
- & -           & $0$ & $2.59$ & $0.505$ & $0.473$  & $0.505$ \\
3 & Triangle    & $2$ & $2.55$ & $1.236$ & $1.250$  & $2.442$ \\
- & -           & $0$ & $2.52$ & $1.178$ & $1.250$  & $2.267$ \\
4 & Tetrahedron & $2$ & $2.63$ & $1.648$ & $1.675$  & $2.884$ \\
- & -           & $0$ & $2.63$ & $1.584$ & $1.654$  & $2.626$ \\
4 & Square      & $2$ & $2.52$ & $1.445$ & $1.485$  & $2.071$ \\
- & -           & $0$ & $2.51$ & $1.416$ & $1.234$  & $1.954$ \\
- & Bulk        & $0$ & $2.81$ & $3.671$ & $3.704$  & - \\
\hline

\end{tabular}
\caption{\label{tab:cluster}
	Geometries, magnetic moment $M$, average bond length $d$,
	binding energy $E_b$ and dissociation energy $E_d$
	of the Pd$_n$ clusters for $n=2-4$.
	The corresponding value of Pd bulk is also given for comparison.
	}
\end{table}

In general, the average bond length of Pd cluster at $M=2$ $\mu_B$
is significantly smaller than the bulk value ($\sim2.8$ {\AA}).
As $n$ increases, $d$ and $E_b$ increase accordingly because the average
number of dangling bonds per atom decreases.
For the $n=4$ case,
though the tetrahedral structure has significantly lower energy ($0.81$ eV) in gas phase,
the square structure becomes important when the underlying symmetry of the TiO$_2$ surface
is taken into account.
The dissociation energy of Pd$_3$ and Pd$_4$ are much larger than that of Pd$_2$,
suggesting in gas phase the dimer is much easier to break into separated entities (monomers).

For the low energy spin-triplet states,
we found the highest occupied molecular orbital (HOMO)
of Pd$_3$ is degenerate and partially occupied.
In this case,
introducing the Jahn-Teller distortion\cite{Jahn1937} can break the symmetry,
and further lower the energy.
Hence these obtained triplet states are not the true ground states.
As pointed out by Futschek \textit{et~al.},\cite{Futschek2005}
the GGA functional is not ensemble v-representable\cite{Ullrich2001}
and hence incapable of describing this effect correctly.\cite{Balasubramanian1989}
However, we argue that the introduction of the surface-cluster interaction
can effectively break the symmetry and remove such degeneracy,
leaving only the gas phase results affected.
This gives a constant shift in the calculated adsorption energies.
No qualitative change of our main conclusions is expected.

\subsection{Pd$_n$ on the stoichiometric surface}\label{subsec:sto}
%Geometries and energetics
The stoichiometric TiO$_2$ is an insulator with a band-gap of $3$ eV.
Its (110) surface can be easily prepared in UHV conditions.\cite{Diebold2003}
As shown in Fig.~\ref{fig:notion},
the surface can be viewed as a composition of alternating rows of
Ti$_{5f}$ atoms and the bridging O$_b$ atoms,
both of which have dangling bonds owing to the surface cleavage.

\begin{figure}[tb]
	\begin{center}
		\includegraphics[width=0.7\textwidth]{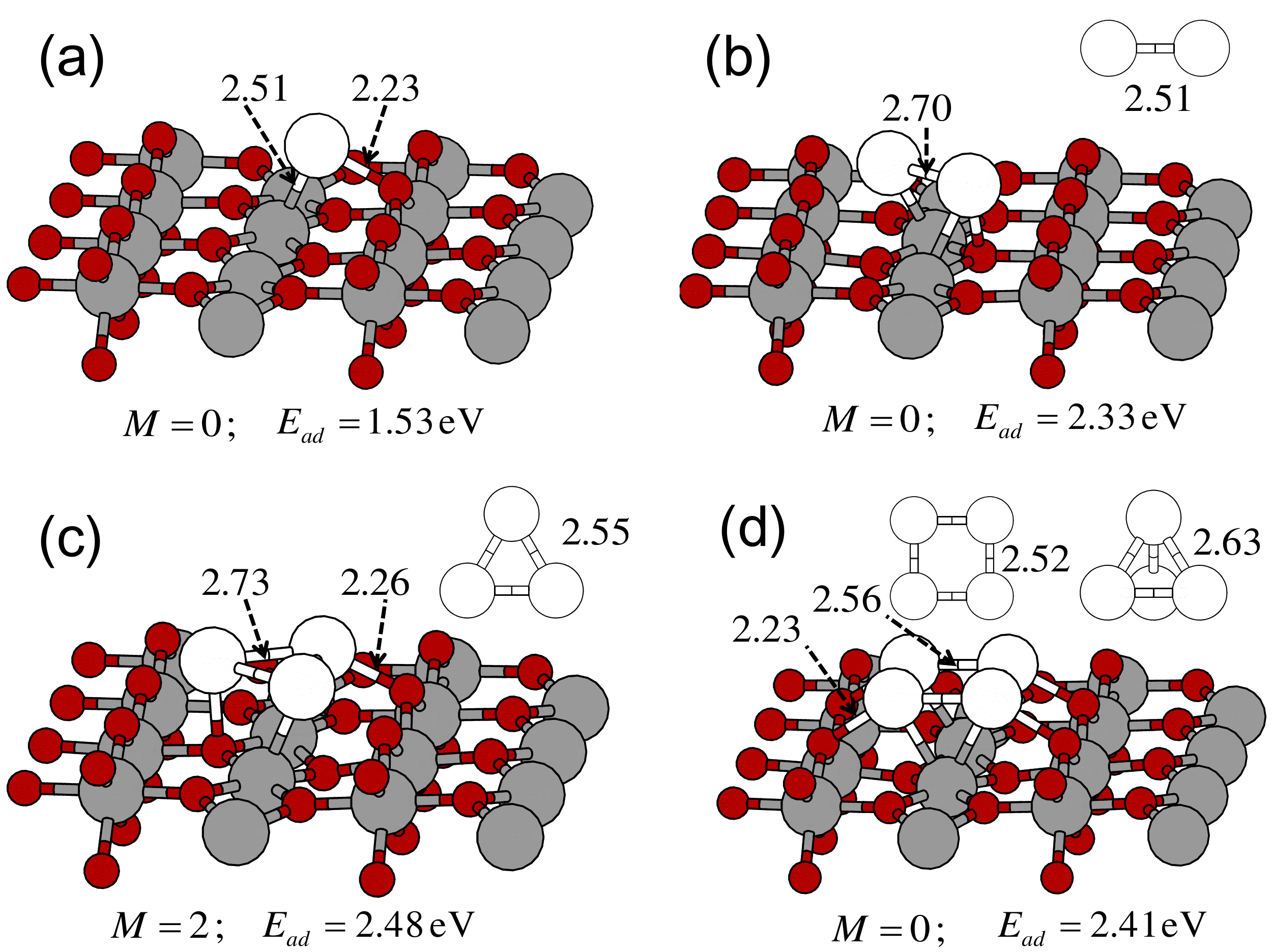}
	\end{center}
	\caption{\label{fig:sto}
	Adsorption geometries of Pd$_n$ cluster for $n=1-4$ on
	the stoichiometric TiO$_2$ (110) surface.
	The corresponding magnetic moment $M$ and adsorption energies
	$E_{ad}$ are given below the geometries.
	Selective bond lengths (in {\AA}) of Pd--Pd and Pd--O bond are given with arrows.
	Insets: ground state geometries of Pd$_n$ cluster in gas phase.
	In the case of Pd$_4$, the ground state is the tetrahedron whereas
	the square planar structures has energy $0.81$ eV higher.}
\end{figure}

We compute the adsorption energy as the difference of DFT total energies
of the two components (the cluster and the surface) before and after the adsorption:
\begin{equation*}
	E_{ad}[n] = E[\mbox{Surf}] + E[\mbox{Pd}_n] - E[\mbox{Surf+Pd}_n],
%\label{eqn:Ead}
\end{equation*}
which is always positive by definition.
The sintering energy $E_s[n]$ has a similar definition of the dissociation energy in gas phase
(see Sec.~\ref{subsec:gas}):
\begin{equation*}
	E_{s}[n] = E[\mbox{Surf+Pd}_{n-1}] + E[\mbox{Surf+Pd}_1]
		- E[\mbox{Surf+Pd}_n] - E[\mbox{Surf}].
%\label{eqn:Es}
\end{equation*}
$E_s$ measures how easily a Pd$_n$ cluster can be decomposed into a Pd$_{n-1}$
and a Pd monomer on the surface.
The adsorption geometries for Pd$_n$ with the highest adsorption energies are shown
in Fig.~\ref{fig:sto}.
The adsorption energies, sintering energies, magnetic moment
and average transferred electrons of selected atoms
(including Pd and the nearby O$_b$ atom bonding to Pd)
are listed in Tab.~\ref{tab:sto}.

\begin{table}[tb]
	\begin{tabular}{ccccccc}
		\hline
		n & Geometry & $M$ ($\mu_B$) & $E_{ad}$(eV) & $E_s$(eV) & $\Delta Q$[Pd](e/atom)
			& $\Delta Q$[O$_b$](e/atom)\\
		\hline
		1 & Monomer     & $0$ & $1.53$ &  --     &  $-0.24$  & $0.07$ \\
		2 & Dimer       & $0$ & $2.33$ &  $0.54$ &  $-0.16$  & $0.06$ \\
		3 & Triangle    & $2$ & $2.48$ &  $1.06$ &  $-0.13$  & $0.05$ \\
		4 & Square      & $0$ & $2.41$ &  $1.28$ &  $-0.11$  & $0.06$ \\
		4 & Tetrahedron & $2$ & $2.40$ &  $1.27$ &  $-0.08$  & $0.05$ \\
		\hline

	\end{tabular}
	\caption{\label{tab:sto}
	Magnetic moment, adsorption energy $E_{ad}$, sintering energy $E_s$,
	and electron transfer $\Delta Q$ of Pd$_n$ ($n=1-4$) clusters adsorbed on
	stoichiometric TiO$_2$ surface.
	The minus sign in $\Delta Q$ means a decrease in the number of electrons
	on the atom.}
\end{table}

The studies of the adsorption properties of Pd clusters on stoichiometric TiO$_2$
surfaces\cite{Bredow1999,Kawazoe2006,Sanz2007}
are not as abundant as those of other transition metal clusters
on the titania surfaces.
The Pd monomer and dimer were investigated by Bredow and Pacchioni,
who showed that the preferred binding site for Pd$_1$
is between two O$_b$ atoms with calculated adsorption energy $1.0$ eV.\cite{Bredow1999}
Sanz and M\'arquez argued that allowing relaxation of the surface gives the h$_o$
as the preferred adsorption site of the Pd monomer,
which is inclining towards a nearby O$_b$ atom and forms a right angle
with Ti$_{5f}$ below.\cite{Sanz2007,Sanz2000}
Our results confirms Sanz and M\'arquez's conclusion (see Fig.~\ref{fig:sto}~(a)).
We found that adsorption of the Pd monomer on the h$_o$ site
is $0.3$ eV lower in energy than that between two O$_b$ atoms.

The Pd dimer has several configurations of adsorption in a small energy range.
One possible way is to adsorb on two adjacent h$_o$ sites along the O$_b$ row,
as discovered by Sanz and M\`arquez.
In this case,
the distance between the two Pd atoms in the dimer is stretched
from $2.51$ {\AA} to $2.75$ {\AA}, to commensurate with the underlying TiO$_2$ lattice
(the lattice constant is $3.06$ {\AA} along the (001) direction).
Another configuration, which is $0.12$ eV lower in energy,
is the Pd dimer bridging two rows of O$_b$ atoms with an angle of about $120$ degrees,
as shown in Fig.~\ref{fig:sto}~(b).
Both Pd atoms are still on the h$_o$ sites.
The sintering energy, $E_s$,
which is the energy cost of breaking the dimer into two monomers, is $0.54$ eV.
This value is slightly lower than that in the gas phase ($0.63$ eV),
suggesting that the Pd dimer is relatively stable on the stoichiometric titania surface.

For the Pd trimer and tetramer,
one notices that because of the cluster-surface interaction,
the adsorbed structures are all \textit{in-plane planar}.
However, for the Pd$_4$ cluster in gas phase,
we found that the tetrahedron Pd$_4$ is $0.81$ eV lower in energy
than the square Pd$_4$ isomer,
whereas the two isomers have almost identical energies when adsorbed on titania.
The square structure is slightly more favored (by $0.01$ eV).
The adsorption geometry of the square structure is shown in Fig.~\ref{fig:sto}~(d).
The incommensurability between the surface structure and the ground state cluster geometry
in the gas phase forces the Pd--Pd bond length to be stretched by about $6$\% on average.
The sintering energies of Pd$_3$ and Pd$_4$ are lower compared to their gas phase values
owing to the cluster-surface interaction.
The tendency of wetting the surface is quite different from the case of
the noble metals, such as Au and Pt,
the clusters of which prefer to grow along the surface normal,
leading to either 3D structures,
or out-of-plane ``standing'' geometries.\cite{Hammer2009,Heyden2010}
The in-plane adsorption picture confirms
the experimental observation of the Anderson group,\cite{Kaden2009}
of which the authors conclude that for $n<10$ the adsorbed Pd clusters are mostly in-plane 2D.
The hollow site h$_o$ is consistently the most preferred adsorption site for the Pd atoms,
as long as the geometry of the cluster allows.

%Magnetic property
In general, we found that the energy differences between the spin-singlet and
the spin-triplet states are in the range of $0.2$--$0.3$ eV,
though their relative order may interchange compared to the gas phase results.
For example, the cluster-surface interaction quenches
the high-spin state of the Pd dimer,
whereas the spin-triplet states of the Pd trimer survives.
%Charge transfer
To characterize the oxidation state of the adsorbed Pd$_n$ clusters,
we performed Bader charge analysis to quantify
the amount of charge transfer between the surface and the cluster,
which is shown in Tab.~\ref{tab:sto}.
The average number of electrons on the surface O$_b$ atom before the adsorption reads $7.04$.
We see that %unlike the case of Pd/Al$_2$O$_3$ system,\cite{Sanz2002a,Sanz2002b,Sanz2001}
there is no significant charge transfer between the cluster and the surface.
The bridging oxygen atom which forms bonds with Pd cluster gains $0.06$e on average.
Compared to the charges in gas phase,
Pd$_1$ donates about $0.24$e to the surface,
while for Pd$_2$--Pd$_4$ these values increases to $0.32$e, $0.39$e and $0.44$e respectively.
The transferred charges are not evenly distributed within the cluster.
In general, this anisotropic distribution of charges could influence the surface chemistry,
in particular, the adsorption and activation of certain gas molecules.
This aspect will be further explored in later publications.

\begin{figure}[tb]
	\begin{center}
		\includegraphics[width=0.7\textwidth]{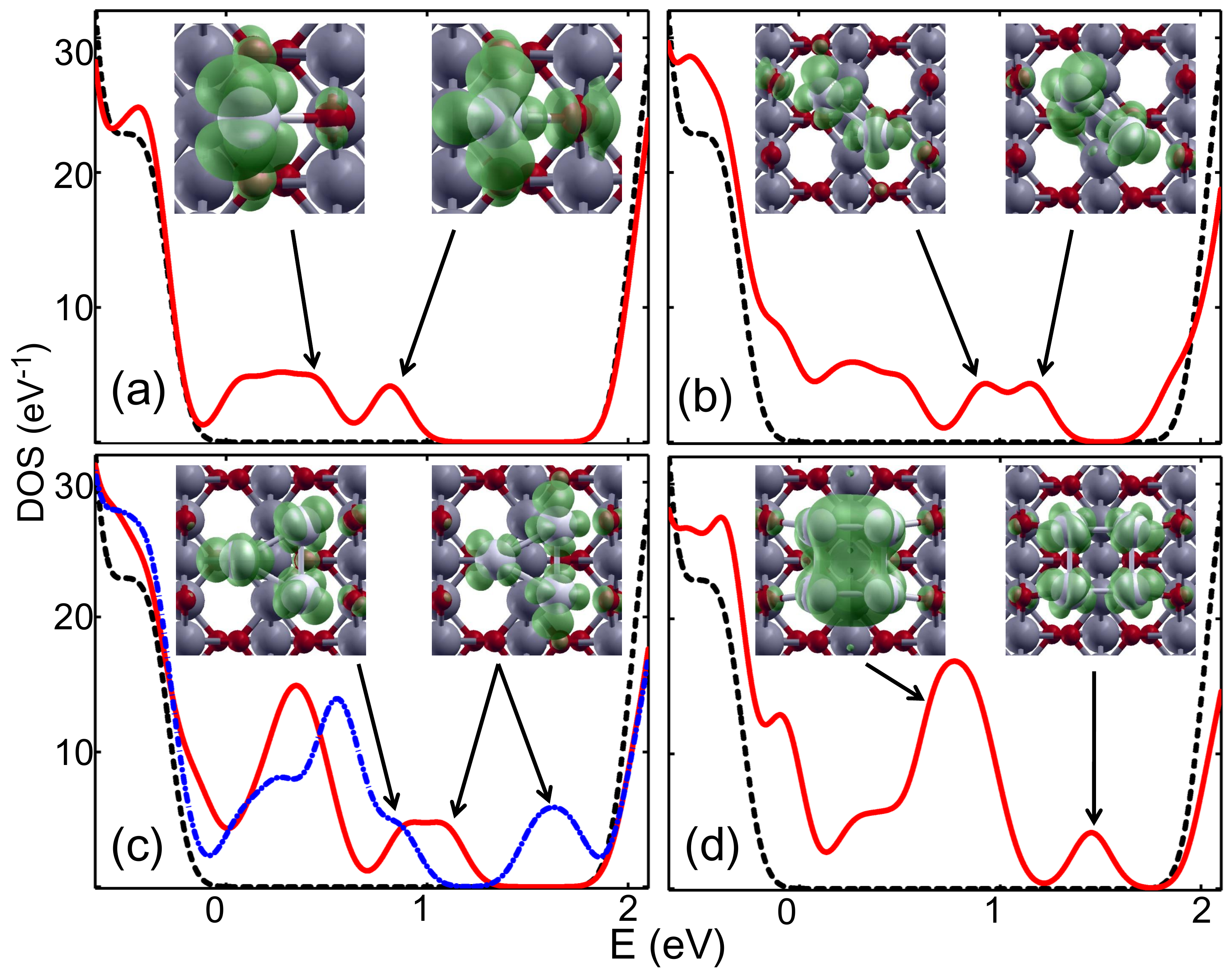}
	\end{center}
	\caption{\label{fig:StoDOS}
	Density of states (DOS) of the adsorbed Pd$_n$ clusters ($n=1-4$)
	on stoichiometric titania surface.
	The DOS of the pure TiO$_2$ surface is plotted as black dashed line.
	For each $n$, selected gap-states are shown in the insets.
	The cut off of the isosurface is chosen to be $0.001$ e/bohr$^3$.
	(The cut off is fixed throughout the paper unless stated otherwise.)
	The arrows mark their corresponding positions on the DOS.
	}
\end{figure}

%Electronic structures and bonding analysis
The small charge transfer indicates that Pd clusters are only slightly oxidized,
largely retaining their gas phase chemistry.
To further understand the nature of the surface-cluster interaction in terms of molecular bonding,
we computed the density of states (DOS) of the supercell and plotted
selected Pd states inside the TiO$_2$ band-gap,
shown in Fig.~\ref{fig:StoDOS}.
For Pd$_1$,
the original degenerate 4d states now split into five states spanning
an energy window about $1.3$ eV.
Though these gap states largely retain the $d$-character,
orbital hybridizations with nearby O$_s$ and O$_b$ atoms are visible.
%Particularly, the HOMO indicates a strong bond with the bridging oxygen atom,
%as concluded from the molecular binding geometries.
For $n=1-4$, all HOMO states suggest weak bonds are formed between
the Pd atoms and the nearby surface O$_s$ and O$_b$ atoms.
By performing the so-called constrained space orbital variation method,\cite{Sanz2002a}
Sanz and M\`arquez argued that the surface-cluster interaction comes largely
from orbital polarization of Pd,\cite{Sanz2007}
which can be seen from the spatial distribution of the HOMOs.
%However, from the pictures shown in Fig.~\ref{fig:StoDOS}~(a),
%a simple polarization of d-orbital is not convincing enough to generate the HOMO state.
The relative height difference in the DOS of the band-gap states comes from
the symmetry requirement imposed by the underlying TiO$_2$ surface.
For the $n=4$ case,
the adsorption of square-shaped cluster is slightly preferred
over that of the tetrahedral one,
because the planar cluster has a large number of states
with four-fold symmetry imposed by the surface (shown as a high peak in the DOS),
as the inset of Fig.~\ref{fig:StoDOS}~(d) shows.
This effect makes the planar cluster having the energy comparable
to that of the tetrahedral structure,
completely closing the gap of $0.81$ eV between
the ground state energies of the two geometries in gas phase.

\subsection{Pd$_n$ on the surface with an oxygen vacancy}\label{subsec:oxvac}
Surface vacancies are the most common point defects seen on TiO$_2$ surface.
When the bridging oxygen is removed,
a pair of electrons is left to fill the nearest available states.
Experimentally,
the spectrum corresponding to Ti$^{3+}$ states were observed by
X-ray photo spectroscopy (XPS) when oxygen vacancies were created,
suggesting the existence of localized Ti$3d$ state from the reduction
of the electron lone pair.
Traditional DFT methods\cite{Vanderbilt1994,Lindan1997,Hammer2004}
failed to capture this phenomenon owing to
the well known self-interaction errors\cite{PerdewZunger1981}
built in the approximate functionals.
Using the hybrid B3LYP functional,
Di Valentin \textit{et~al.} successfully obtained the gap states
for systems with oxygen vacancies.\cite{DiValentin2006}
The DOS shows an energy split of about $0.1$ eV in the TiO$_2$ band-gap.
Within the framework of DFT+$U$,
Morgan and Watson showed that there are two different regimes
describing the localization of electrons.\cite{Morgan2007}
For $U<4.2$ eV, electrons partially localize on several subsurface Ti ions.
A qualitative change was seen when $U$ was increased above $4.2$ eV.

\begin{figure}[tb]
	\begin{center}
		\includegraphics[width=0.7\textwidth]{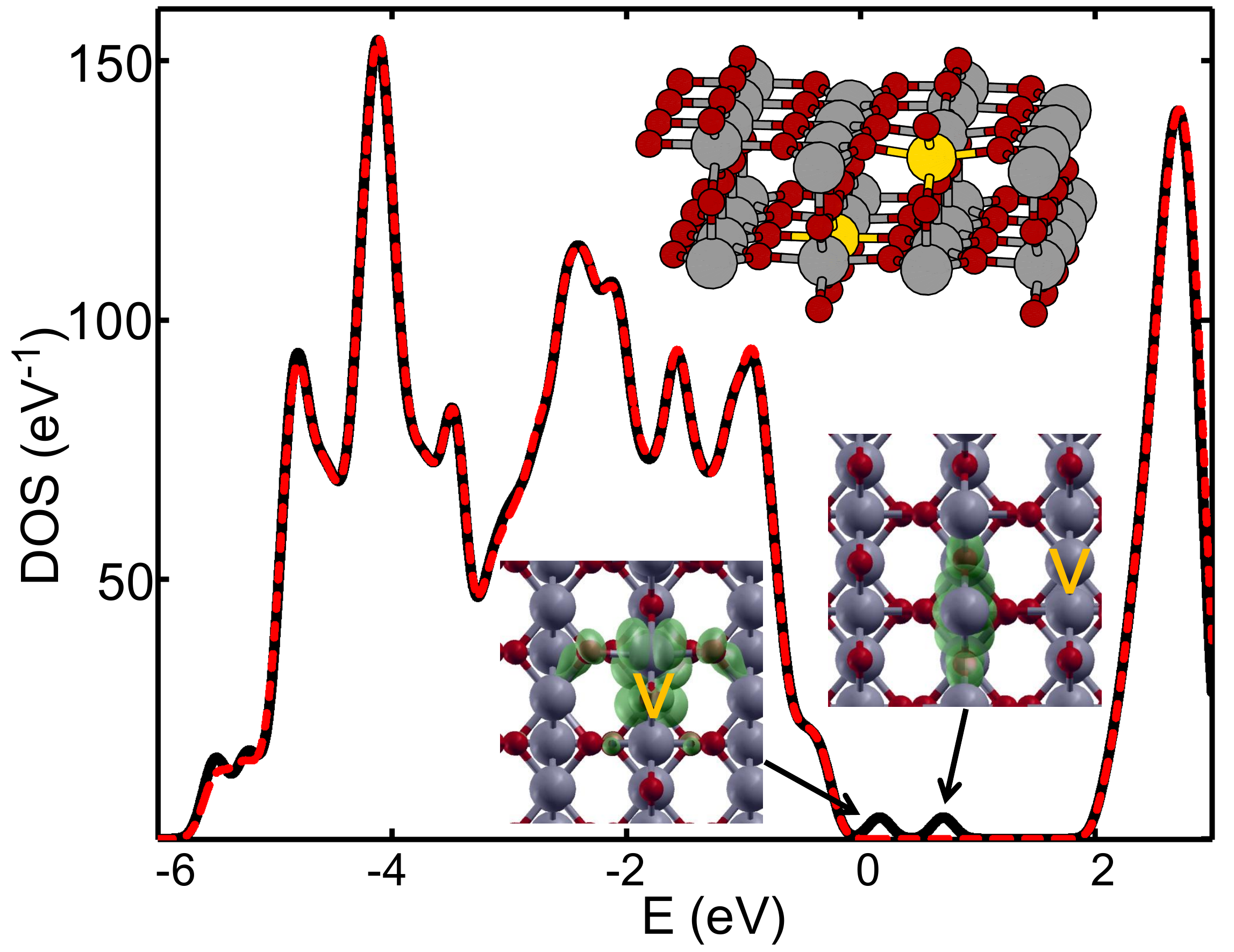}
	\end{center}
	\caption{\label{fig:OVPd0_dos}
	DOS of the defective TiO$_2$ surfaces with an oxygen vacancy.
	The inset shows the structure of the corresponding surface.
	The position of the localized electron is shown in golden color.
	The spin-resolved DOS are shown as solid black
	and dashed red lines for spin up and down respectively.
	The two gap states with localized electrons
	are shown in the insets with arrows marking their positions
	on the DOS.
	}
\end{figure}

With the self-consistently calculated $U$ ($4.2$ eV) on all Ti atoms,
we found that multiple sites of localization can be achieved,
agreeing with Deskins \textit{et~al.}'s conclusion.\cite{Dupuis2011}
The lowest energy configuration involves one electron localizing on one Ti$_{6f}$
neighboring the vacancy and one on a next nearest neighbor subsurface Ti ion
under the surface Ti$_{5f}$ row.
The spin-triplet ground state has splitted DOS in the titania band-gap,
as shown in Fig.~\ref{fig:OVPd0_dos}.
The localization picture is similar to Calzado and co-workers' results.\cite{Calzado2008}
Owing to the inequivalent localization sites,
both energies and orbitals of the two gap-states differ from each other.
The electron localizing near the surface vacancy is lower in energy (by $0.6$ eV).

The reason that various localization sites were obtained roots deeply in the use of
the parametric approach of correcting the self-interaction error of density functionals.
Different choices on the values of $U$ or the percentage of the Hatree-Fock
exchange included in the hybrid functional determines not only the energy of the gap states,
but also the positions of the localized electrons.\cite{Selloni2008}
(see Ref.~\onlinecite{Dohnalek2010} by Dohn\'alek and co-workers for a complete review.)
Experimentally, based on the resonant photoelectron diffraction measurement,
Kruger \textit{et~al.} found that the excessive charges
are largely shared by several surface and subsurface Ti sites with
a dominant contribution from the subsurface Ti ion\cite{Kruger2008},
which is in line with our calculation (see Fig.~\ref{fig:OVPd0_dos}).
On the other hand, using the low temperature STM technique,
Minato \textit{et~al.} showed that the two excessive electrons do not localize
directly on the vacancy but on neighboring surface Ti$_{5f}$ ions.\cite{Minato2009}
In reality, the dynamics of the surface structures at non-zero temperatures
may play an important role in the electronic localization.
Combining DFT+$U$ and the Car-Parrinello molecular dynamics,
Kowalski and co-workers have shown that the localized electrons are in fact very mobile,
and ``hop'' between different sites because of
the thermal vibration of the TiO$_2$ lattice.\cite{Kowalski2010}
Both splitted and non-splitted DOS were obtained
depending on the transient state of the localized electrons.
The subsurface layer was found to be occupied by excessive electrons about $70$\% of time,
with the surface layer and the third trilayer populated at $20$\% and $10$\% of time respectively.
This result suggests that most of previous works (including this paper) actually captured snapshots
of this complicated phenomenon at zero temperature.
It would be very interesting to see
how the adsorbates interact with the localized electrons near the oxygen vacancy
at finite temperature.
%The surface conductivity (needed by the STM) neary the oxygen vacancy is polaronic in nature.
%For applications only concerning with the position of
%gap states rather than the electron dynamics,
%we assume that using the lowest energy configuration can
%give qualitatively correct results.
In the following we show that,
even at zero temperature,
the adsorption of additional molecules near the oxygen vacancy can affect the local geometry
and hence change the energetics and electron localization sites in various ways.

\begin{table}[tb]
	\begin{tabular}{cccccc}
		\hline
		n & Geometry & $M$ ($\mu_B$) & $E_{ad}$(eV) & $E_s$(eV) & $\Delta Q$[Pd](e/atom) \\
		\hline
		1 & Monomer     & $2$ & $1.79$ &  --     &  $-0.28$  \\
		2 & Dimer       & $0$ & $2.36$ &  $0.04$ &  $0.26$  \\
		3 & Triangle    & $0$ & $1.99$ &  $0.02$ &  $0.13$  \\
		4 & Tetrahedron & $0$ & $1.37$ &  $0.22$ &  $0.10$  \\
		\hline

	\end{tabular}
	\caption{\label{tab:oxvac}
	Magnetic moment, adsorption energy $E_{ad}$, sintering energy $E_s$,
	and electron transfer per Pd atom $\Delta Q$ of Pd$_n$ ($n=1-4$) clusters adsorbed
	on the TiO$_2$ surface next to an oxygen vacancy.}
\end{table}

\begin{figure}[tb]
	\begin{center}
		\includegraphics[width=0.7\textwidth]{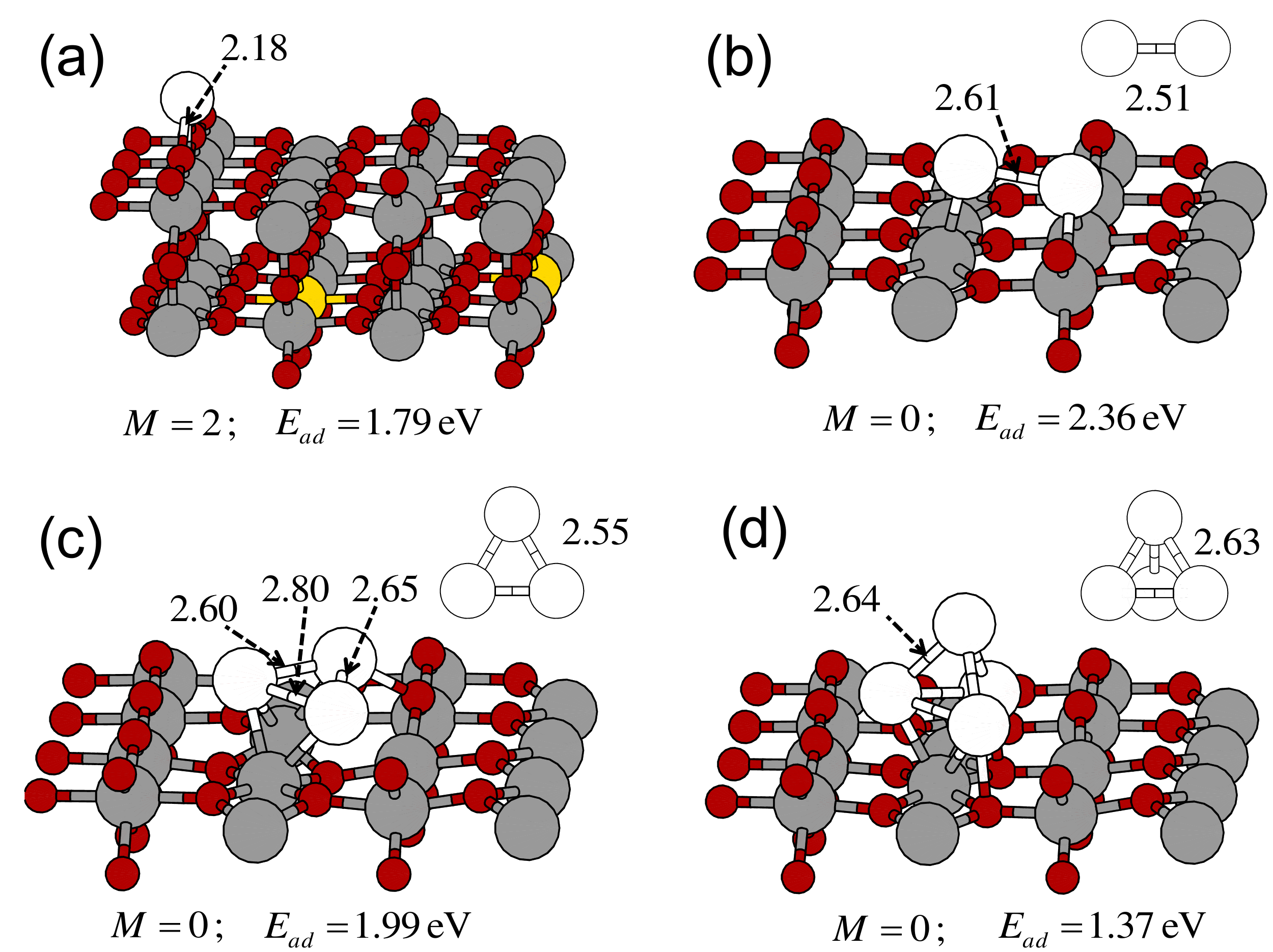}
	\end{center}
	\caption{\label{fig:oxvac}
	Adsorption geometries of Pd$_n$ cluster for $n=1-4$ nearby an oxygen vacancy.
	The corresponding magnetic moment $M$ and adsorption energies
	$E_{ad}$ are given below the geometries.
	Selective bond lengths (in {\AA}) of Pd--Pd and Pd--O bonds are shown with arrows.
	For the Pd monomer in (a), the electron localization sites are shown in golden color.
	The gas phase cluster structures are shown in the insets.
	}
\end{figure}

Fig.~\ref{fig:oxvac}~(a--d) shows the adsorption geometries and energetics of a Pd$_n$ cluster
close to a surface oxygen vacancy.
The magnetic moment and sintering energies are listed in Tab.~\ref{tab:oxvac}.
In Sec.~\ref{subsec:sto}, we show that on the stoichiometric surface,
Pd has a tendency to binding to protruding O$_b$ atoms.
This picture is further confirmed in the adsorption nearby a vacancy.
Contrary to the results of Ref.~\onlinecite{Sanz2007},
the preferred adsorption site for Pd$_1$ is not on the vacancy but away from it.
And the ``usual'' h$_o$ site next to an O$_b$ atom now is less favored (by $0.25$ eV)
as compared to the bridging site between two O$_b$ atoms (see Fig.~\ref{fig:oxvac}~(a)).
The total energy gain by moving away from the vacancy to the bridging position between
two O$_b$ atoms is $0.65$ eV.
The adsorption energy of the geometry shown in Fig.~\ref{fig:oxvac}~(a) is $0.26$ eV
higher than that of the Pd monomer on stoichiometric surface.
The presence of the Pd monomer also affects the localization site of the two excessive surface electrons
--- both of which now localize in the subsurface layer,
in an effort to maximize their mutual distance and those with the Pd monomer.
Such local repulsion effect cannot be captured without
correcting the self-interaction error in the density functional.

Similar to the results in Sec.~\ref{subsec:sto},
the Pd dimer has multiple adsorption geometries within the energy range of $0.2$ eV.
Fig.~\ref{fig:oxvac}~(b) shows the configuration with the lowest energy.
the Pd dimer shows no preference over the type of surface ---
either stoichiometric or with an oxygen vacancy,
the adsorption energies are roughly the same.
The original four-fold symmetry enforced by the surface is
broken because of the presence of the vacancy.
For $n=3$ and $4$, the consequences of lifting the four-fold surface symmetry is that
the trimer is no longer an equilateral but a scalene triangle lying on the surface;
the adsorption of the square structure now takes (instead of releasing) $0.67$ eV.
The square configuration is disfavored by more than $2$ eV,
compared to the tetrahedral configuration.
In all cases, the sintering energy of Pd cluster next to an oxygen vacancy is lowered by
$0.5-1.0$ eV as compared to those on the stoichiometric surface.
Particularly, $E_s$ is close to zero for the Pd dimer and trimer.
It can be concluded that the oxygen vacancy does not act as nucleation site
but repels and dissociates Pd clusters if deposited directly on top of the vacancy.

%Magnetic property
For the pure surface with an oxygen vacancy,
strong electron localization is observed,
and the spin-triplet states is $1.6$ eV lower in energy than the spin-singlet.
When attached with Pd clusters,
the situation changes dramatically.
Except the case of Pd$_1$,
where the electronic character of the surface vacancy is largely preserved
because of the non-direct Pd adsorption,
in all other cases the ferromagnetic coupling between the two localized electrons is attenuated.
The localization on individual Ti ions is completely quenched because of
the surface-cluster interaction.
In fact, if the Pd monomer is moved directly above the vacancy site,
similar attenuation of surface spin states is achieved.

%Charge transfer
Except for the case of Pd$_1$,
the Bader charges in Tab.~\ref{tab:oxvac} show that the usual ``electron-donor'' role
played by the Pd cluster is now reversed.
Overall the surface vacancy is back donating about $0.4$e to $0.5$e
to the cluster --- a significant amount of charge to influence possible catalytic reactions
taking place on top of the cluster.
The charge flow explains why the Pd cluster doesn't bind strongly
to the underlying Ti atoms on the vacancy or stays near
compared to the stoichiometric case.
From Tab.~\ref{tab:sto} we know that Pd cluster gets stabilized by donating part of
its electrons to the surface.
In the case of the Pd monomer bound to
the bridging site shown in Fig.~\ref{fig:oxvac},
the charge transfer from the Pd atom to the surface is $0.28$e.
However we found that when the monomer is moved closer to an oxygen vacancy,
electrons flow in the opposite direction.
For h$_o$ sites, the charge accumulation on the Pd monomer can be as large as $0.50$e,
indicating that the localized electrons on the Ti ions have a strong tendency
of reducing Pd,
and effectively decreases the energy benefit of the Pd cluster
forming a bond with the underlying surface.
Binding to atoms with high electron affinity can reduce such energy penalty,
leading to the bridging site between two O$_b$ atoms as the most stable adsorption site for the Pd monomer.

\begin{figure}[tb]
	\begin{center}
		\includegraphics[width=0.7\textwidth]{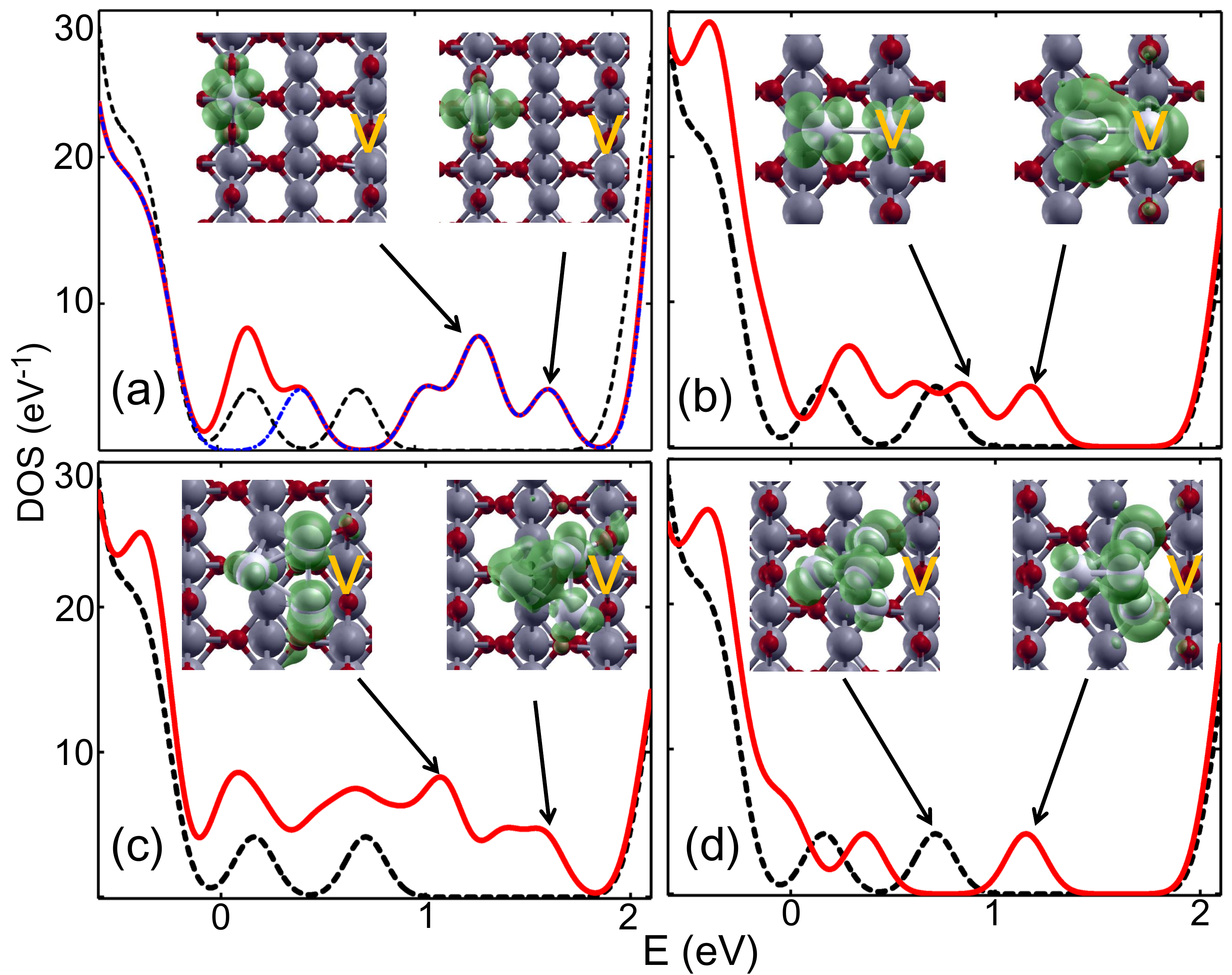}
	\end{center}
	\caption{\label{fig:OVDOS}
	DOS of the adsorbed Pd$_n$ clusters ($n=1-4$) close to an oxygen vacancy.
	The DOS of the pure defective surface is plotted in black dashed lines for comparison purpose.
	The spin-resolved DOS (up and down) with the adsorbates
	are shown as red solid and blue dash-dotted lines respectively.
	Insets: selected states with position on the DOS marked by arrows.
	The position of the oxygen vacancy is marked with ``V''.
	}
\end{figure}

%Electronic structures and bonding analysis
The DOS of the adsorbed Pd$_n$ clusters and selected gap states are shown in Fig.~\ref{fig:OVDOS}.
It can be seen that the gap states have almost no overlap with nearby oxygen atoms.
Unlike the case of adsorption on the stoichiometric surface,
the back-donation of electrons from the surface preserves the electronic structure
of the cluster in gas phase.
These additional charges also made the overall system favor the spin-singlet state.

\subsection{Pd$_n$ on the surface with a Ti-interstitial atom}\label{subsec:tiint}
\begin{figure}[tb]
	\begin{center}
		\includegraphics[width=0.7\textwidth]{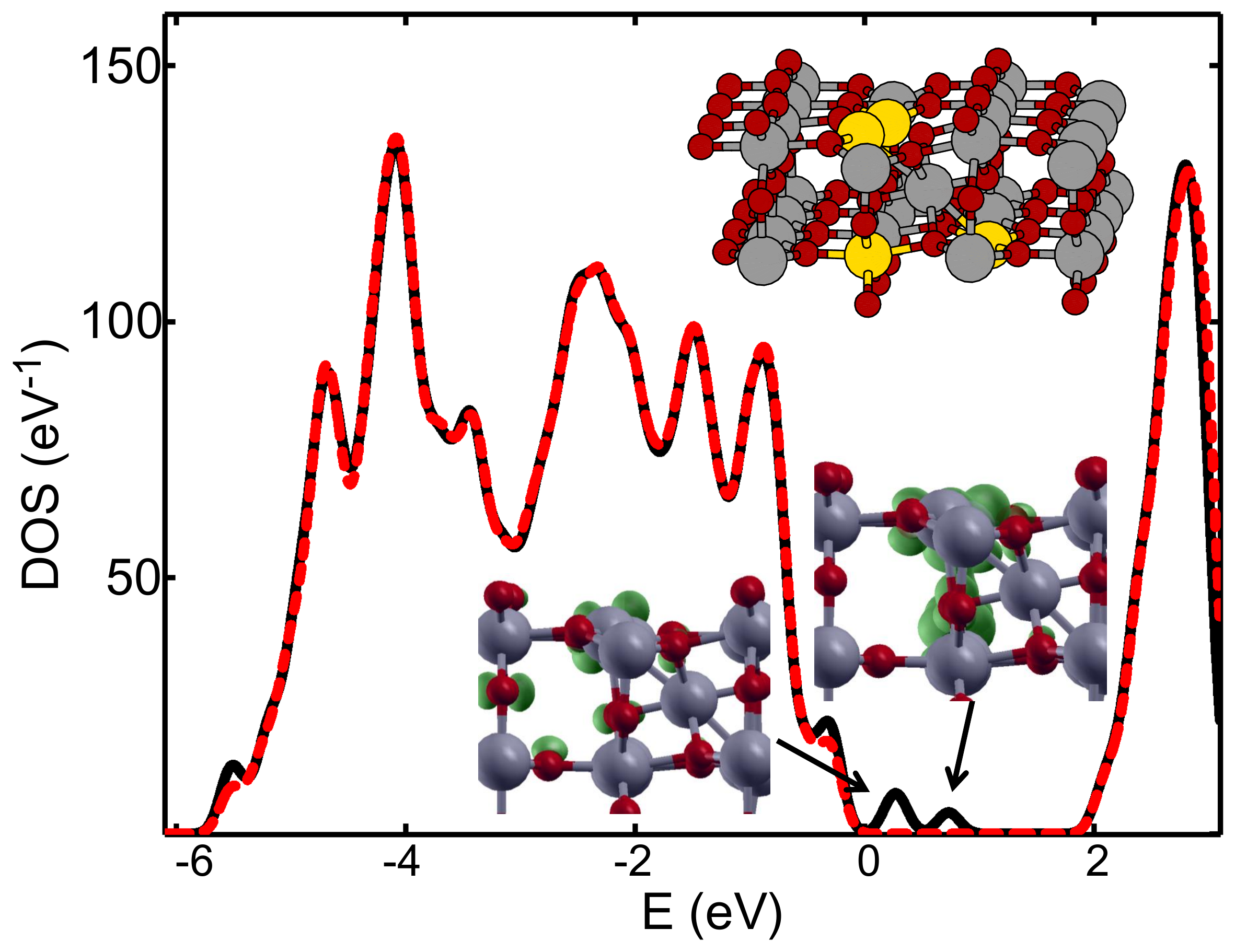}
	\end{center}
	\caption{\label{fig:IntPd0_dos}
	DOS of the TiO$_2$ surfaces with a subsurface Ti-interstitial atom.
	The inset shows the geometry of the surface model.
	Ti ions with localized electrons are highlighted in golden color.
	The spin resolved DOS are shown as solid black
	and dashed red lines for spin up and down respectively.
	Selected gap states formed by localized electrons
	are shown in the insets with arrows marking their positions on the DOS.
	}
\end{figure}

Conventionally, the oxygen vacancy was believed to be largely responsible
for the observed gap states (with some contributions
from surface hydroxyls\cite{Suzuki2000,Kowalski2009,DiValentin2006,Hammer2010}).
However, Wendt \textit{et~al.} pointed out that under certain experimental conditions,
to a large extent the surface chemistry on reduced titania is associated with
the Ti-interstitials existing in the near surface region.\cite{Wendt2008}
The additional Ti atom may be located at various depths.\cite{Wendt2008,Hammer2009,Bennett2010}
Through interstitial exchange mechanism,
these Ti-interstitial atoms can diffuse from the subsurface to the surface layer.
To avoid potential artifacts associated with our computational setup
where two bottom trilayers of TiO$_2$ were fixed to their bulk positions,
we only consider the case that a Ti-interstitial is located in the first trilayer,
as shown in Fig.~\ref{fig:IntPd0_dos}.

The interstitial Ti atom reduces the surface with its four valence electrons,
two of which are found to localize on the surface Ti$_{5f}$ atoms next to the interstitial atom.
The other two localize in the subsurface layer.
The magnetic coupling between these electrons is relatively weak,
giving an energy difference of only $0.01$ eV between the spin-triplet ($M=2~\mu_B$)
and the spin-quintet ($M=4~\mu_B$) states.
The spin-triplet is slightly lower in energy.
When a Pd cluster is attached to the surface,
a preference over the spin-quintet was identified.
In what follows,
to facilitate direct comparisons of the electronic structures
of clusters of different sizes on the equal footing,
we show only the results of spin-quintet states.
The magnetization and the qualitative picture of localization
confirms the one obtained by Mulheran \textit{et~al.},
who studied the diffusion of Ti interstitials
with DFT+$U$ and an atomistic potential method.\cite{Bennett2010}
However, we obtained splitted DOS with the HOMO placed $1.2$ eV below the CBM,
whereas in their paper all Ti$3d$ states share the same energy,
whose distance to the CBM is less than $0.5$ eV.
The HOMO and HOMO-1 are shown in the insets of Fig.~\ref{fig:IntPd0_dos}.
As one may see,
though strongly localized in nature,
the excess electrons are, in fact, shared by more than one Ti ions.

\begin{figure}[tb]
	\begin{center}
		\includegraphics[width=0.7\textwidth]{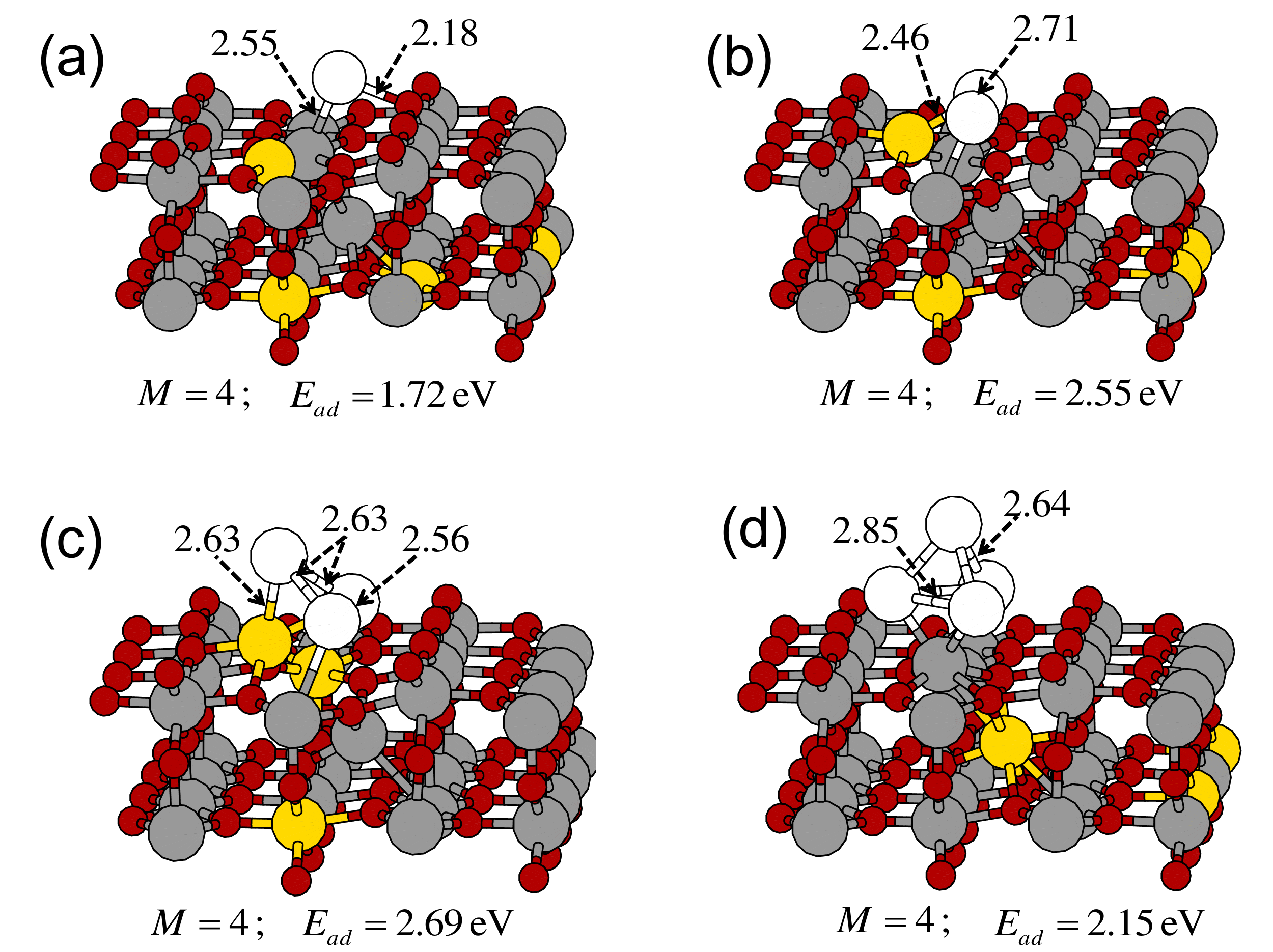}
	\end{center}
	\caption{\label{fig:int}
	Adsorption geometries of Pd$_n$ cluster for $n=1-4$ with an Ti-interstitial.
	The magnetic moment $M$ and adsorption energies
	$E_{ad}$ are given below the geometries.
	Selective bond lengths (in {\AA}) between Pd--Pd and Pd--O are shown with arrows.
	}
\end{figure}

To the best of our knowledge,
the adsorption of Pd$_n$ clusters ($n=1-4$) on the titania (110) surface with
a subsurface Ti-interstitial atom has not been investigated yet.
The obtained geometries and adsorption energies are shown in Fig.~\ref{fig:int}~(a--d).
In addition, the Bader charges and sintering energies are given in Tab.~\ref{tab:int}.
The Pd monomer, again, strongly prefers to bind to the O$_b$ atoms.
The adsorption geometry is very similar to the stoichiometric result (see Fig.~\ref{fig:sto}~(a)),
though the adsorption energy $E_{ad}$ is $0.19$ eV higher.
Contrary to the case of oxygen vacancy,
the presence of subsurface Ti-interstitial atoms may attract Pd monomers
and act as nucleation centers for cluster growth.

For the Pd dimer, similar to the other cases,
multiple adsorption geometries with small energy differences were identified.
The lowest energy configuration is shown in Fig.~\ref{fig:int}~(b).
The corresponding spin-triplet state has a similar geometry but higher in energy by $0.58$ eV.
Compared to other types of surface, the subsurface Ti-interstitial atom stabilizes
the adsorption of the Pd dimer by $0.2$ eV.
The surface structure is strongly distorted owing to the adsorption of the Pd$_2$ cluster.
One Ti$_{5f}$ atom between the cluster and the interstitial atom
is forced to move out of the surface layer and form a bond with an adjacent O$_b$ atom.
This type of surface reconstruction is very similar to the minimum energy pathways
obtained in the simulated diffusion of Ti interstitial atoms
from subsurface to surface layer.\cite{Wendt2008,Bennett2010}
%The adsorption of the Pd dimer may accelerate the growth of TiO$_2$ island
%by lowering the energy barrier of Ti diffusion.

Pd$_3$ and Pd$_4$ show a similar tendency of binding to the protruding surface layer Ti$_{5f}$ atom
(see Fig.~\ref{fig:int}~(c--d)).
For Pd$_3$, we are unable to reach self-consistency for the spin-triplet state,
whereas the spin-singlet state has the same energy as does the spin-quintet.
In the case of Pd$_4$, the situation is reversed,
the calculation could not reach self-consistency for the spin-singlet.
But the energy of the spin-triplet state is almost identical to
that of the spin-quintet one.
Due to the presence of the Ti-interstitial atom,
the four-fold surface symmetry is effectively lifted.
The binding geometries are no longer in-plane 2D but
show a tendency of growing along the surface normal.

\begin{table}[tb]
	\begin{tabular}{cccccc}
		\hline
		n & Geometry & $M$ ($\mu_B$) & $E_{ad}$(eV) & $E_s$(eV) & $\Delta Q$[Pd](e/atom) \\
		\hline
		1 & Monomer     & $4$ & $1.72$ &  --     &  $-0.20$  \\
		2 & Dimer       & $4$ & $2.55$ &  $0.38$ &  $-0.01$  \\
		3 & Triangle    & $4$ & $2.69$ &  $0.86$ &  $0.11$  \\
		4 & Tetrahedron & $4$ & $2.15$ &  $0.63$ &  $0.00$  \\
		\hline

	\end{tabular}
	\caption{\label{tab:int}
	Magnetic moment, adsorption energy $E_{ad}$, sintering energy $E_s$,
	and electron transfer $\Delta Q$ of Pd$_n$ ($n=1-4$) clusters adsorbed
	on TiO$_2$ surface with a subsurface Ti interstitial atom.}
\end{table}

%Charge transfer
The average Bader charge per Pd atom are shown in Tab.~\ref{tab:int}.
The charge transfer behavior shows an interesting sign change as $n$ increases.
The Pd monomer,
unlike the case of oxygen vacancy,
is an electron donor rather than an acceptor.
The Pd trimer is slightly negatively charged.
The Pd dimer and tetramer have unequally distributed charges over their atomic components,
with an average charge transfer close to zero.
Overall the charge transfer between the cluster and the surface is not significant.
This may be related to the fact that both Pd and the Ti-interstitial atom are competing with
each other to donate electrons to the surface.

\begin{figure}[tb]
	\begin{center}
		\includegraphics[width=0.7\textwidth]{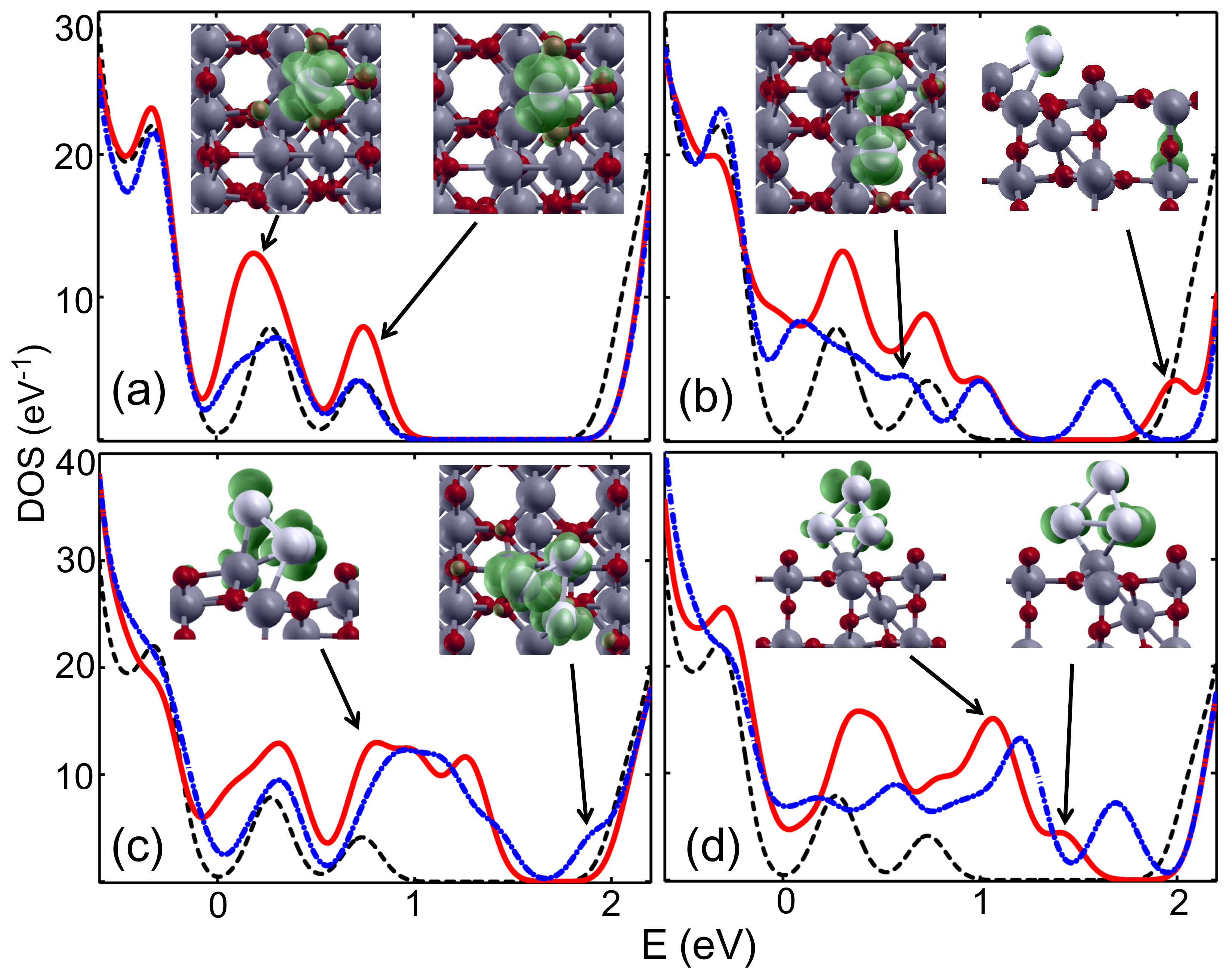}
	\end{center}
	\caption{\label{fig:IntDOS}
	DOS of the defective TiO$_2$ surfaces with subsurface Ti-interstitial atom.
	The inset shows the structure of the corresponding surface.
	The spin resolved DOS are shown as solid red
	and dashed-dotted blue lines for spin up and down respectively.
	The DOS of the pure surface is plotted in black dashed line.
	The two gap states with localized electrons
	are shown in the insets with arrows marking their positions
	on the DOS.
	}
\end{figure}

%Electronic structures and bonding analysis
The spin-resolved DOS of the adsorbed Pd$_n$ cluster and selected gap states
are depicted in Fig.~\ref{fig:IntDOS}.
Similar to the case of oxygen vacancy,
there is no significant bonds forming between the cluster and the surface.
Most gap states have similar character of Pd cluster states in gas phase,
with its directions polarized by the underlying surface.
For the case of the Pd monomer and dimer,
the high magnetic moment is contributed largely by the surface.
We obtain four Ti ions with one of its $3d$ states nearly filled by an electron.
From the DOS, one can see that the difference between the DOS
of the up spin and down spin nearly coincides with the pure surface DOS.
When $n=3$ and $4$,
the cluster-surface interaction partially lift the electron localization---
only three electrons are found localized on Ti ions.
The remaining $1\mu_B$ contribution to the magnetic moment has no clear spatial origin.

\subsection{Cluster growth and sintering}\label{subsec:sinter}
Clusters deposited on top of metal-oxides surfaces
may absorb or lose a number of atoms and change their sizes at non-zero temperatures.
Such sintering effects often lead to annihilation of small-sized clusters,
which might be crucial for catalyzing desired reactions.\cite{Goodman1995,HenrichCox1996}

The traditional theory on cluster sintering suggests that
most cluster nucleation and growth processes involve the
exchange of monomers between different clusters.\cite{Schmidt1996}
It is assumed that monomers are much more mobile than larger clusters.
Using scanning tunneling microscopy (STM),
Xu \textit{et~al.} investigated
the Pd cluster nucleation and growth on the titania surface experimentally.\cite{Xu1997}
At low coverage ($0.012$ ML),
they found that Pd clusters grow in 2D,
with structures pseudomorphic to the substrate TiO$_2$ surface.
In their experiment, only the Pd dimer, trimer and larger clusters were observed,
and no existence of the Pd monomer was confirmed.
On the other hand, Kaden \textit{et~al.} have measured the
catalytic activity of CO oxidation of size-selected
Pd cluster deposited on the titania surface.
Clear differences in the catalytic activities of Pd$_1$ and Pd$_2$ were observed,
suggesting the Pd monomer can exist on the surface at UHV condition.

\begin{figure}[tb]
	\begin{center}
		\includegraphics[width=0.7\textwidth]{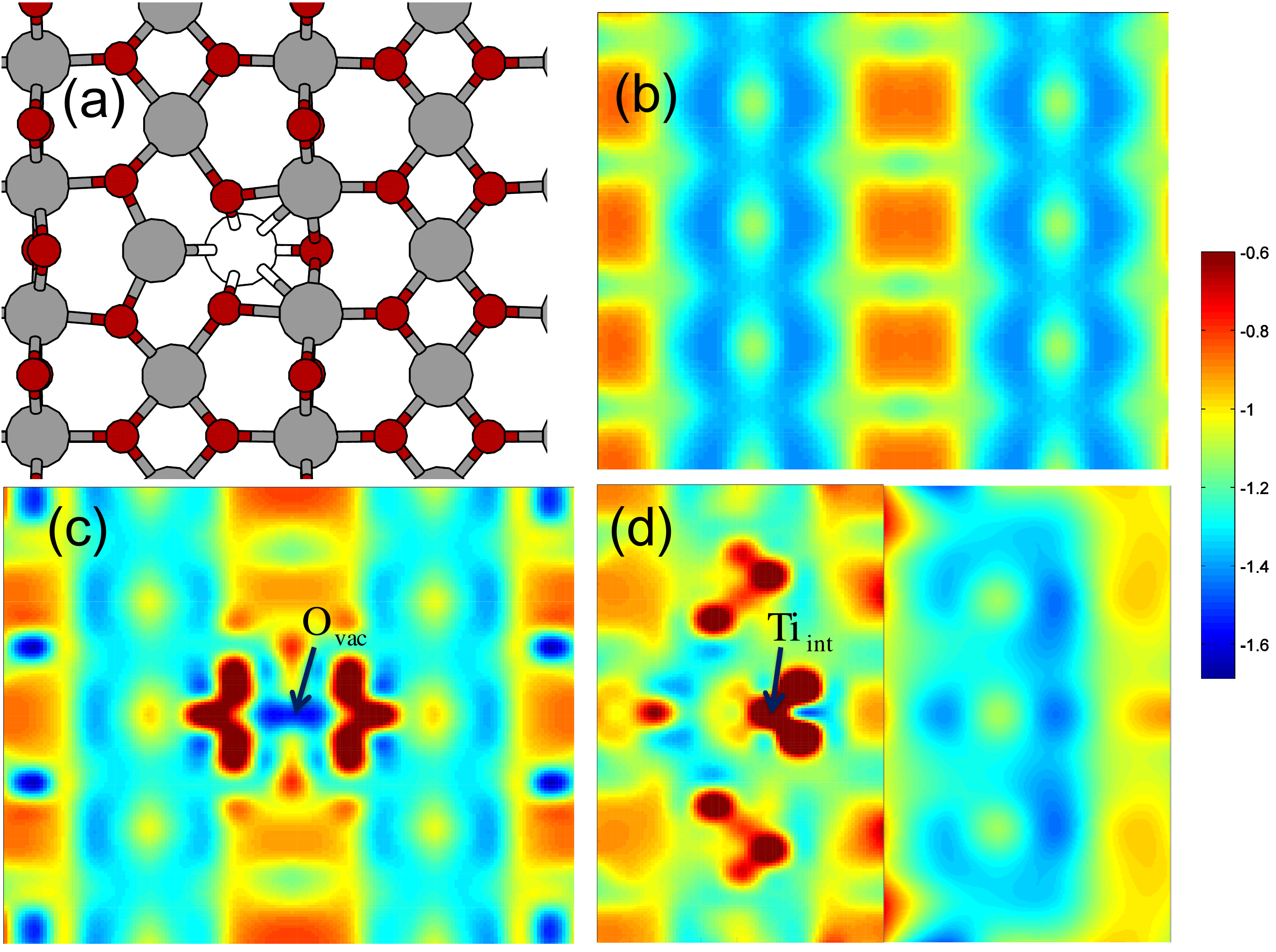}
	\end{center}
	\caption{\label{fig:PES}
	Potential energy surfaces (PES) for the Pd monomer on the
	stoichiometric and defective surfaces.
	(a) Top view of the surface cell (with a Ti-interstitial atom as an example)
	aligned with the PES graphs in (b--d).
	(b) PES on the stoichiometric surface.
	(c) PES on the surface next to an oxygen vacancy.
	(d) PES on the surface with a Ti-interstitial atom.
	(N.B.,
	the discontinuity of (d) is due to technical reasons when the PES was generated.)
	The positions of the vacancy and the Ti-interstitial are marked by arrows.
	The color bar shows the corresponding energy: $-E_{ad}$.
	For the dark red regions in (c) and (d),
	the energies are positive and exceed the value denoted by the color bar.
	}
\end{figure}

To quantify the mobility of the Pd monomer,
we calculated PESs of a Pd atom adsorbed on the stoichiometric and the defective titania.
On the stoichiometric surface,
we confirm that the Pd monomer is extremely mobile along the O$_b$ rows,
forming a ``two-way traffic'' between two O$_b$ rows with the h$_o$ sites as energy minima.
The PES is shown in Fig.~\ref{fig:PES}~(b).
The energy barrier between energy minima is only $0.05$ eV.
To cross the O$_b$ row,
the energy minimum path involves
the Pd atom first moving to the site between two adjacent O$_b$ atoms
then further moving to the other side.
The highest barrier along the path is $0.3$ eV,
which is accessible at the room temperature.
When the Pd monomer moves nearby an oxygen vacancy,
the h$_o$ sites are still the local energy minima,
however the bridging sites between two O$_b$ atoms one row away from the vacancy
are more favored by the Pd monomer,
as discussed in Sec.~\ref{subsec:oxvac}.
The vacancy itself is very difficult to reach (energy barrier larger than $1$ eV),
though there is an energy minimum right on the vacancy once the barrier has been overcome
(see Fig.~\ref{fig:PES}~(c)).
In the case that a Pd atom is close to a subsurface Ti-interstitial atom,
the PES is more distorted by the presence of the additional Ti atom.
We found a small spot for the Pd atom to bind to the underlying interstitial atom,
with surrounding high energy barriers.
The PES rapidly approaches the stoichiometric one when Pd moves away from
the Ti-interstitial atom, as shown in the right half section of Fig.~\ref{fig:PES}~(d).

To simulate the thermodynamics of cluster growth and sintering effect,
we used the PES of the Pd monomer on the three types of surfaces,
and construct a 2D square lattice model describing the movement of the Pd monomers
with the Monte Carlo method.
The lattice sites are chosen to be a repeating two rows of h$_o$ sites
and one row of bridging sites between two O$_b$ atoms.
The Pd atom is allowed to jump to the nearest and the next nearest neighbors.
The hopping rates are modified using the harmonic transition state theory,\cite{Masel2001}
in which the barriers are read from the saddle points between the adsorption sites on the PES.
We assume the classical cluster sintering model,
in which only the Pd monomer is mobile.
The larger clusters grow or diminish by acquiring/losing the Pd monomers.
Each time we randomly pick one Pd atom and move it according to the Metropolis algorithm.
If one Pd atom collides into another Pd$_n$ cluster,
a Pd$_{n+1}$ cluster is formed with probability one.
If the chosen atom is already in a cluster,
an energy penalty of the corresponding sintering energy $E_s$ is added to determine
whether that atom will leave the cluster.
We simulate a $16\times16$ super-lattice with the periodic boundary condition.
Each cell of the lattice contains one TiO$_2$ ($4\times2$) unit,
and is assumed to have a random type of surface.
The PESs of the adjacent cells are assumed to be independent of their neighboring cells.
To simulate typical experimental conditions,
we chose $10$\% cells to have an oxygen vacancy and
$5$\% cells with a Ti-interstitial atom.
The remaining cells have the stoichiometric surface.
The Pd coverage is fixed at $0.16$ ML ($\sim410$ atoms) throughout.
MC simulations were done over $1000$ initial configurations with randomly distributed Pd monomers.
For each configuration, a multiple of $10^9$ MC runs at fixed temperatures
were executed to ensure the system reaches its equilibrium distribution.
Finally, our calculation is essentially zero-temperature based.
Using molecular dynamics simulations,
Miguel \textit{et~al.} studied the stability of Pd clusters 
on the titania surface, and found that 
the surface-cluster interaction decreases appreciably
for temperatures higher than $800$ K.\cite{Sanz2007a,Negra2003}
Deposited clusters may adopt gas phase geometries 
(e.g. the tetramer may acquire a tetrahedral structure), 
and become metallic at higher temperatures. 
Such effects are not taken into account in our modeling.

\begin{figure}[tb]
	\begin{center}
		\includegraphics[width=0.7\textwidth]{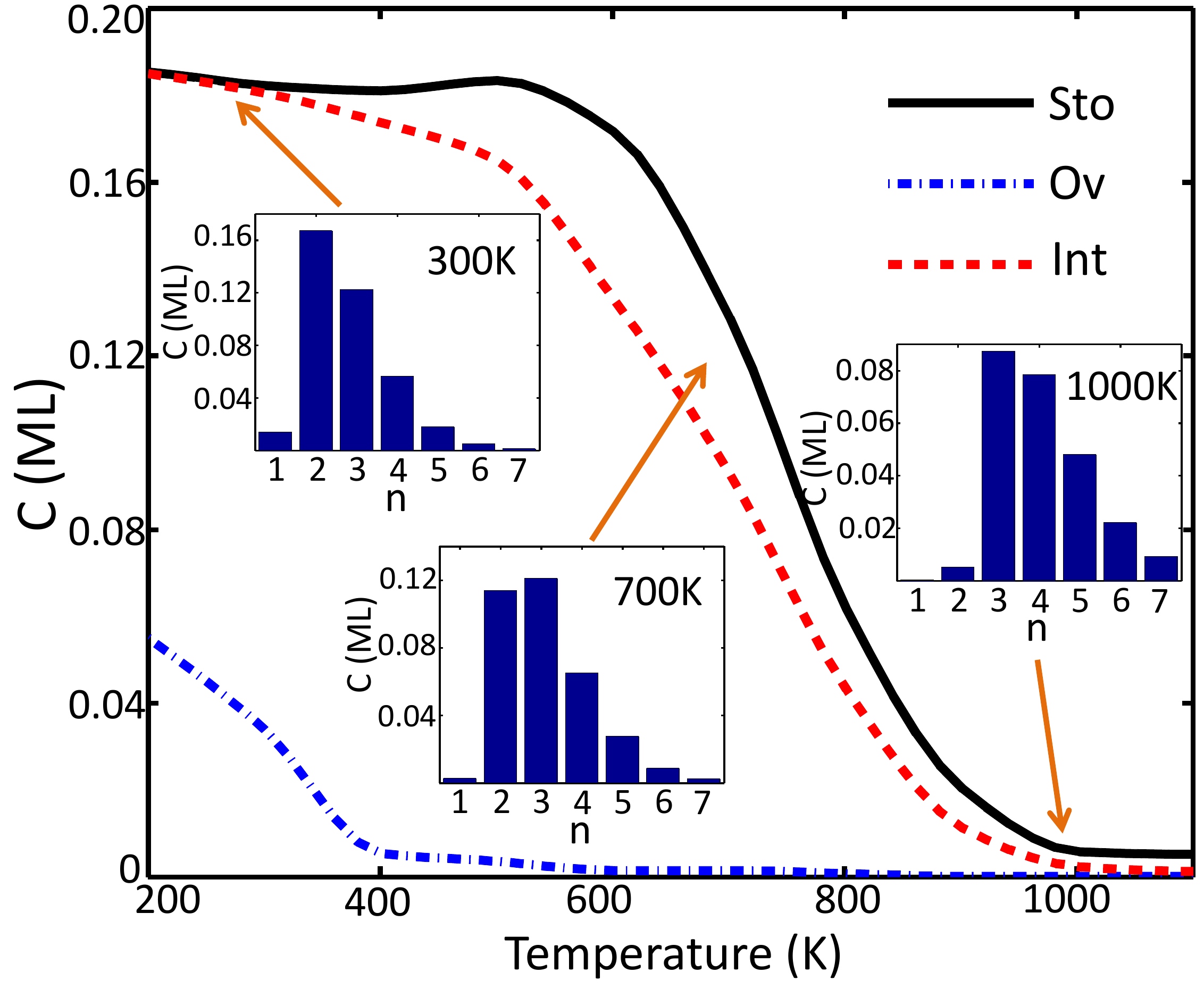}
	\end{center}
	\caption{\label{fig:MC}
	Density $C$ (in ML) of the Pd dimer (Pd$_2$) as a function
	of temperature on different types of surfaces.
	The black solid line (Sto) shows the one on the stoichiometric surface.
	The blue dashed dotted line (Ov) gives the
	density on the surface with an oxygen vacancy,
	while the red dashed line (Int) corresponds to the surface
	with a Ti-interstitial atom.
	Three insets shows the density distribution of Pd$_n$ cluster
	with respect to cluster size $n$ at temperature
	$300$ K, $700$ K, and $1000$ K.
	The MC simulations were performed at $0.16$ ML the Pd monomer coverage.
	}
\end{figure}

The density changes of the Pd dimer as a function of temperature
on different types of surfaces are drawn in Fig.~\ref{fig:MC}.
The thermal stability of clusters of different sizes is generally determined
by the relative strength of their sintering energies $E_s$.
At all temperatures,
there is only a trace amount of the Pd monomers remaining at equilibrium,
since the sintering energy $E_s$ of the Pd monomer is implied to be zero.
At low temperatures,
the most abundant cluster is the Pd dimer and,
Pd$_n$ clusters are generally immobile.
The Pd dimer is very stable on the stoichiometric surface
as well as close to a Ti-interstitial atom for temperatures below $600$ K.
When the temperature reaches above $400$ K,
the possibility of finding a Pd dimer near an oxygen vacancy is practically zero.
We found that this is also true for larger clusters,
in line with the small sintering energies on the surface with an oxygen vacancy
(see Tab.~\ref{tab:oxvac}).
The sintering process in which the Pd dimers lose individual Pd atoms
and exchanging these to grow into larger clusters begins to take effect at $600$ K.
The maximum of the size distribution moves to larger clusters,
which can be told by comparing the insets of Fig.~\ref{fig:MC} at different temperatures.
The $E_s$ of the Pd trimer is about twice that of the Pd dimer
on the stoichiometric surface and near a Ti-interstitial.
This implies that the trimer should be stable below $1200$ K,
which is partly confirmed by the size distribution diagram at $1000$ K.

\section{Conclusion and Acknowledgement}
We have performed a systematic analysis on the geometric
and electronic structures of small Pd clusters on
the stoichiometric and two defective TiO$_2$ (110) surfaces.
We find that unlike the noble metals, which prefer binding to the surface oxygen vacancies,
the Pd clusters tend to adsorb on the stoichiometric surface
or next to a subsurface Ti-interstitial atom.
Owing to the surface-cluster interaction,
Pd clusters have 2D in-plane structures when adsorbed on the stoichiometric surface.
For defective surfaces,
we demonstrate that DFT+$U$ is capable of reproducing the experimentally
observed gap-states and predicts that the localized electrons and the high-spin states
of the surface may be quenched when a Pd cluster is adsorbed.
We show that the Pd monomer is very mobile on the TiO$_2$ surface
and contributes greatly to the cluster sintering effect.

The support form the ACS PRF grant 51052-DNI6 is acknowledged.
Most calculations were performed on the UCLA Hoffman2 shared cluster.
A portion of the research was performed using EMSL,
a national scientific user facility sponsored by the Department of Energy's Office
of Biological and Environmental Research and located at Pacific Northwest National Laboratory.
We thank Professor Scott Anderson (University of Utah) for a helpful discussion.

%\bibliographystyle{jcp}
%\bibliography{citation}

\end{document}